\documentstyle{laa}

\begin{document}
\thesaurus{07  
	   (07.09.1; 
	   )}

\title{Perihelion Concentration of Comets \\
       I. Discussion of the Published Methods}

\author{ Jozef Kla\v{c}ka }
\institute{Astronomical Institute,
Faculty for Mathematics and Physics,
Comenius University, \\
Mlynsk\'{a} dolina,
842 15 Bratislava,
Slovak Republic }
\date{}
\maketitle

\begin{abstract}
The problem of (non)random distribution of points on the sphere is investigated.
Published procedures for obtaining preferred direction and preferred plane
for points on the sphere (in the sky) are discussed. It is shown that
the published methods are incorrect, and, as a consequence, the results
obtained by these methods cannot be considered to be significant.
\end{abstract}

\section{Introduction}
Neslu\v{s}an (1995, 1996) and other authors (see references in
Neslu\v{s}an 1996), e. g., Tyror 1957, Yabushita 1979,
Bogart and Noerdlinger 1982
have discussed the distribution of perihelia
of long-period comets in the sky. The aim of this paper
is to discuss the mathematical treatment of the methods used by these authors.

\section{Perihelion Point Preferred Direction}
Perihelion point of a long-period comet in the sky may be characterized
by direction cosines $l_{i}$, $m_{i}$, $n_{i}$. By definition of
direction cosines, the relation
$l_{i}^{2} ~+~m_{i}^{2} ~+~n_{i}^{2} =$ 1 holds.

If we have $N$ comets (perihelion points in the sky), we have a set
of $N$ triples of direction cosines:
$\{ l_{i} \} _{i=1}^{N}$, $\{ m_{i} \} _{i=1}^{N}$, $\{ n_{i} \} _{i=1}^{N}$
(subscript $i$ denotes the $i$-th comet).

Neslu\v{s}an (1995, 1996) (and other authors -- see references in
Neslu\v{s}an 1996) finds the preferred direction of the points of perihelia
and the result is presented by Eqs. (1)-(3) in
Neslu\v{s}an (1996):
\begin{equation}\label{1}
< l > ~ = \left ( \sum_{i=1}^{N} w_{i} ~ l_{i} \right ) ~/~
	\left ( \sum_{i=1}^{N} w_{i}  \right )
\end{equation}
\begin{equation}\label{2}
< m > ~ = \left ( \sum_{i=1}^{N} w_{i} ~ m_{i} \right ) ~/~
	\left ( \sum_{i=1}^{N} w_{i}  \right )
\end{equation}
\begin{equation}\label{3}
< n > ~ = \left ( \sum_{i=1}^{N} w_{i} ~ n_{i} \right ) ~/~
	\left ( \sum_{i=1}^{N} w_{i}  \right ) ~,
\end{equation}
where $w_{i}$ is the weight of the $i-$th comet in the given calculation.
It is supposed that some weights may differ from unity in Neslu\v{s}an's (1996)
paper, which is a generalization of the preceding papers. However, none of
the authors present any argument why one should use Eqs. (1)-(3)
(with $w_{i} =$ 1 for $i =$ 1 to $N$, or, $w_{i} \ne w_{j}$ in general).
They could be
correct only if $\{ l_{i} \}_{i=1}^{N}$,$\{ m_{i} \}_{i=1}^{N}$,
$\{ n_{i} \}_{i=1}^{N}$ would be independent quantities
(the sample mean -- the expected value).
Moreover, it is only the author's decision if he calculates the preferred perihelion
direction for comets discovered from a given hemisphere with weight
$w =$ 1/2 for comets discovered independently from both hemispheres
(the same holds for the older papers $w_{i} =$ 1 for $i =$ 1 to $N$).

The statement made by Neslu\v{s}an is that Eqs. (1)-(3) determine direction
cosines $< l >$, $< m >$ and $< n >$ of the preferred direction.
This statement is used also in Eq. (5) in
Neslu\v{s}an (1996), where it is supposed that
$< l >$, $< m >$ and $< n >$ form components of a unit vector (and, the result
of Eq. (5) is used in Eq. (4) in Neslu\v{s}an 1996).

The error is evident. Quantities $< l >$, $< m >$ and $< n >$ do not form
components of a unit vector!
They are not direction cosines of any unit vector
(of a preferred direction).
Moreover, there is no suggestion how to calculate errors for the quantities
presented by Eqs. (1)-(3).

\section{Other Comments on the Preferred Direction}
According to Neslu\v{s}an (1996) and other authors, the significance
of the departure of the perihelion directional distribution from random
distribution is characterized by the mean quadratic distance of the directions
of perihelion points from their preferred direction (see Eq. (4) in
Neslu\v{s}an 1996):
\begin{equation}\label{4}
< D^{2} > ~ = \left ( \sum_{i=1}^{N} w_{i} ~ \cos ^{2} \gamma _{i} \right )
	    ~ / ~ \left ( \sum_{i=1}^{N} w_{i} \right ) ~.
\end{equation}
The statement is: ``In the case of the random
distribution, $< D^{2} > =$ 1/3. The departure from randomness is then
characterized by expression 1/3 ~--~ $< D^{2} >$.'' (below Eq. (5) in
Neslu\v{s}an 1996).

Our comment is that $< D^{2} > =$ 1/3 (more correctly, if
$w_{i} = w$, $i =$ 1 to $N$)
holds only for the special type
of random distribution -- for uniform distribution (uniform in spherical
angles).
Moreover, the obtained
results for 1/3 ~--~ $< D^{2} >$ Neslu\v{s}an compares with ``the standard
deviation'', which is not defined in Neslu\v{s}an (1996). It is defined
only in Neslu\v{s}an (1995) (Eq. (7), p. 39). However, it does not seem
to be correct, since for equal weights (the standard case used by
authors up to Neslu\v{s}an's work) it yields identical 0 (zero).

No preferred direction exists for uniform distribution.
Since the observational data yield a preferred direction, it is important
to calculate errors of the investigated quantities.

\section{Preferred Direction -- The Probability Density Function}
Eqs. (6)-(9) in Neslu\v{s}an (1996) define the procedure of comparing the
real distribution of the data with the theoretical $\chi ^{2} -$distribution
(there is not correct sign in the exponent in Eq. (9) in Neslu\v{s}an (1996),
and, Eq. (9) does not represent the distribution function but the
density function).
Unfortunately, the procedure used in Neslu\v{s}an (1996) is incorrect.
Neslu\v{s}an defines by Eq. (8) the quantity which is the argument of the
density function defined by Eq. (9).
(The motivation in Eq. (8) is incorrect, since the random quantities
$l$, $m$, $n$ are not independent -- $l^{2} ~+~ m^{2} ~+~n^{2} =$ 1 --
and their distributions are not of type N(0, 1).)
However, theoretical density function
does not contain the number of the observed elements (points), in reality!
If one would like to use the idea defined by Eqs. (6)-(7), then the correct
argument in the density function should be the corresponding area of the sphere
(e. g., in square degrees). Thus, Figs. (3)-(5) in Neslu\v{s}an (1996)
are of no usefulness -- {\bf it only seems that real distribution
may be approximated by uniform distribution}. However, Figs. 8
in Neslu\v{s}an (1995) show that in reality the distribution is not
exactly uniform -- the effect of galactic potential partially disturbes
the uniformity (Neslu\v{s}an 1995: it seems that directions of perihelion points
of long-period comets avoid galactic plane).

\section{Preferred Direction -- One More Comment}
The section 5 in Neslu\v{s}an (1996) discusses partial compensation
of asymmetry in comet discoveries. It seems to us not wise to make such
a procedure, since we can make whatsoever -- what we want. It is seen
from Eqs. (10)-(11) in Neslu\v{s}an (1996). We have already commented
the definition of (original -- initial) weights. However,
Eqs. (10)-(11) modify the original weights -- the original weights
are again weighted in order to obtain some better weights. What are
the better and the best weights? Those, which would produce uniform
distribution? In any case, any kind of weights yields some preferred
direction (only special cases yield uniform distribution, however, these
special cases cannot be easily found and justified). The author's role is to
explain the obtained result for the existence of the preferred direction --
it is not enough to say that his kind of weights yields different result
from that obtained by the previous authors.

The correct procedure should have an opposite direction:
one must start with the assumption of uniform distribution, and, making some
known theoretical considerations, he comes to some theoretical prediction;
the theoretical prediction must be consistent with the preferred direction
yielded by the observational data.

\section{Significance of the Results}
As we have already mentioned, it is important to calculate errors for
the calculated quantities. Only the correct calculation of errors can
present the significance of the calculated quantities.

Tyror's method (Tyror 1957, used also, e. g., in Bogart and Noerdlinger 1982)
of calculation of the errors for the significance of the determination
of the preferred plane can be summarized in the following equations:
\begin{eqnarray}\label{5}
\lambda _{i} &=& ( \vec{p} \cdot \vec{x_{i}} ) ^{2}  ~~, ~~~
< \lambda > ~ = ~ \sum_{i=1}^{N} \lambda _{i} / N \nonumber \\
\sigma_{< \lambda >} &=&
\sqrt{\sum_{i=1}^{N} ( \lambda _{i} - < \lambda > )^{2} / N / ( N - 1 )} ~,
\end{eqnarray}
where $\vec{x_{i}}$ is unit vector of the perihelion position for the
$i-$th comet, $\vec{p}$ is unit vector normal to the preferred plane.
Supposing normal distribution with the mean $< \lambda >$ and standard
deviation $\sigma_{< \lambda >}$,
Tyror's method compares the value $1/3 ~-~ < \lambda >$ with
$\sigma_{< \lambda >}$.

The first argument against the method of Tyror is trivial. In reality,
unit vector $\vec{p}$ is determined from the observational data
(``measurements''), and, thus, it is not exact quantity. However, its error
is not considered in the calculation of $\sigma_{< \lambda >}$
given by Eq. (5). Thus, real error of $< \lambda >$ is larger than that
given by Tyror's method.

The second argument against the method of Tyror is that although Eq. (5) seems
to be equivalent to calculation of $\sigma_{< \lambda >}$ by the procedure of
calculations of errors, the procedure of Tyror is not standard. Tyror's method
is equivalent to calculation of the error of $\lambda$, calculated by the
standard error-method procedure for measurements, from equation
(eigen-value problem leads to the cubic equation for $< \lambda >$)
\begin{eqnarray}\label{6}
& & \lambda ^{3} - \lambda ^{2} + \left \{ < x^{2} > < y^{2} > ~+~
< x^{2} > < z^{2} > ~+~ \right .
\nonumber  \\
& & \left .
< y^{2} > < z^{2} > ~-~
< x y >^{2} ~-~ < x z >^{2}  ~-~ < y z > ^{2} \right \} \lambda ~+~
\nonumber  \\
& & < x^{2} > < y z >^{2} ~+~ < y^{2} > < x z >^{2} ~+~ < z^{2} > < x y >^{2} ~-~
\nonumber  \\
& & 2~< x y > < y z > < x z > ~-~< x^{2} > < y^{2} > < z^{2} > ~ = ~  0 ~,
\end{eqnarray}
where the errors of quantities $< x^{2} >$, $< y^{2} >$, $< z^{2} >$,
$< x y >$, $< x z >$, $< y z >$ are supposed to be known and are calculated
as, e. g.,
$\sigma_{< x y >}$ $=$
$\sqrt{\sum_{i=1}^{N} ( x_{i} y_{i} - < x y > )^{2} / N / ( N - 1 )}$ ~. \\
Thus, defining
$\lambda_{i} \equiv \lambda$ ($x_{i}^{2}$, $y_{i}^{2}$,
$z_{i}^{2}$, $x_{i} y_{i}$, $x_{i} z_{i}$, $y_{i} z_{i}$),
$< \lambda > = \lambda$ ($< x^{2} >$, $< y^{2} >$,
$< z^{2} >$, $< x y >$, $< x z >$, $< y z >$),
the Tyror's method supposes that the relation
\begin{eqnarray}\label{7}
\lambda_{i} &=& < \lambda > ~+~
\frac{\partial \lambda}{\partial < x^{2} >}
\left ( x_{i}^{2} ~ - ~  < x^{2} > \right ) ~+~
\nonumber \\
& &
\frac{\partial \lambda}{\partial < y^{2} >}
\left ( y_{i}^{2} ~ - ~ < y^{2} > \right ) ~+
\nonumber \\
& &
\frac{\partial \lambda}{\partial < z^{2} >}
\left ( z_{i}^{2} ~ - ~ < z^{2} > \right ) ~+~
\nonumber \\
& &
\frac{\partial \lambda}{\partial < x y >}
\left ( x_{i} y_{i} ~ - ~ < x y > \right ) ~+
\nonumber \\
& &
\frac{\partial \lambda}{\partial < x z >}
\left ( x_{i} z_{i} ~ - ~ < x z > \right ) ~+~
\nonumber \\
& &
\frac{\partial \lambda}{\partial < y z >}
\left ( y_{i} z_{i} ~ - ~ < y z > \right )
\end{eqnarray}
holds, where partial derivatives are calculated on the basis of solution
of Eq. (6), and, higher orders in Taylor expansion of $\lambda_{i}$ are
neglected. Using the right-hand side of Eq. (7), $\sigma_{< \lambda >}$,
defined in Eq. (5), yields the result consistent with that obtained by the
method of Tyror. Eq. (7), in reality, corresponds to the situation when
quantities $x_{i}^{2}$, $y_{i}^{2}$,
$z_{i}^{2}$, $x_{i} y_{i}$, $x_{i} z_{i}$, $y_{i} z_{i}$
are considered to be independently measured (and higher order terms are
negligible, i. e., measured quantities are, say, normally
distributed to a certain degree of approximation).
Of course, this is not true. Thus, the method of Tyror
is not correct.

Finally, we show the third argument that Tyror's method is not correct.
In reality, as it was already mentioned, Eq. (7) is not exact. It would
be not exact even in the case when $x_{i}^{2}$, $y_{i}^{2}$,
$z_{i}^{2}$, $x_{i} y_{i}$, $x_{i} z_{i}$, $y_{i} z_{i}$
would be independently measured. The reason is that higher orders in Taylor
expansion are neglected. In reality, one would not need to consider
an approximation by Taylor expansion since $\lambda_{i}$ could be calculated
from equation analogous to Eq. (6) -- instead of mean values, values of
$x_{i}^{2}$, $y_{i}^{2}$, $z_{i}^{2}$, $x_{i} y_{i}$, $x_{i} z_{i}$,
$y_{i} z_{i}$
should be used. This situation corresponds to the eigen-value problem
\begin{equation}\label{8}
C_{i} ~ \vec{p_{i}} = \lambda_{i} ~ \vec{p_{i}} ~,
\end{equation}
where matrix components are $C_{i11} = x_{i}^{2}$,
$C_{i12} = C_{i21} = x_{i} y_{i}$, $C_{i13} = C_{i31} = x_{i} z_{i}$,
$C_{i22} = y_{i}^{2}$, $C_{i23} = C_{i32} = y_{i} z_{i}$, $C_{i33} = z_{i}^{2}$.
This is, of course, geometric nonsense, since the one point
( $x_{i}$, $y_{i}$, $z_{i}$ ) ( $x_{i}^{2} ~+~y_{i}^{2} ~+~z_{i}^{2} ~=~$ 1 )
cannot determine a plane, which is, according to Eq. (8), characterized by
the plane's normal (unit) vector $\vec{p}$. In accordance with Eq. (8),
one would consider the relevant value $\lambda_{i} =$ 0, and,
\begin{eqnarray}\label{9}
\sigma_{< \lambda >} &=&
\sqrt{\sum_{i=1}^{N} ( \lambda _{i} - < \lambda > )^{2} / N / ( N - 1 )}
\nonumber \\
&=& < \lambda >  / \sqrt{N - 1} ~.
\end{eqnarray}
As an example we mention that the Tyror's result $N =$ 448,
$< \lambda > =$ 0.276 would yield $\sigma_{< \lambda >} =$ 0.013 which
corresponds to the approximate value 0.014 (Tyror's result) obtained on the
basis of Taylor expansion defined by Eq. (7).

We have shown that the results presented in Tyror (1957),
Bogart and Noerdlinger (1982) are of no significance since the method
used in these papers corresponds to construction of a plane when only
one point is known -- this is geometrical nonsense since, in reality,
three points are
needed for constructing a plane.

\section{Concentration of Points and Its Mathematical and Physical Treatment}
Both problems on perihelion concentration of long-period comets -- perihelion
preferred direction and preferred plane for perihelion points -- were solved
by the authors using unit vectors of individual perihelion points. In general,
this method yields different results from those obtained on the basis of the
correct physical treatment of the problem. Physics requires that all problems
must be solved using radii vectors of the perihelion points. One can calculate,
using this way, e. g., the preferred direction of perihelion points based on
the centre of mass of the system. Detail theoretical comparison of both methods --
unit vectors and radii vectors -- will be done in the following paper.

%

\section{Conclusion}
We have shown that mathematical methods used in the problem of determination
of preferred directions and preferred planes for long-period comets are
incorrect. Thus, correct procedures must be elaborated, and, subsequently,
applied to the observational data. The results obtained by such correct methods
must be interpreted, then. All these problems are presented in the following
parts of this series of papers.

\acknowledgements
The author wants to thank to \v{L}. Tur\v{n}a for useful discussion.
Special thanks to the firm ``Pr\'{\i}strojov\'{a} technika, spol. s r. o.''.
This work was also partially supported by Grants VEGA No. 1/4304/97
and No. 1/4303/97.

\end{document}